\documentclass[
reprint,
amsmath,amssymb,
aps,
pra,
]{revtex4-1}

\usepackage{wrapfig}

\usepackage[dvipdfmx]{graphicx}
\usepackage[dvipdfmx]{color}

\usepackage{braket}
\usepackage{dcolumn}
\usepackage{bm}

\newcommand{\naive}{na\"{\i}ve }

\begin{document}

\title{Drag dynamics in one-dimensional Fermi systems}

\author{Jun'ichi Ozaki}\email{Present email address: j.ozaki@aist.go.jp}
\author{Masaki Tezuka}\email{tezuka@scphys.kyoto-u.ac.jp}
\author{Norio Kawakami}\email{norio@scphys.kyoto-u.ac.jp}

\affiliation{Department of Physics, Kyoto University, Kyoto 606-8502, Japan}

\date{\today}

\begin{abstract}
We study drag dynamics of several fermions in a fermion cloud in one-dimensional continuous systems, 
with particular emphasis on the non-trivial quantum many-body effects in systems whose parameters change gradually in real time. 
We adopt the Fermi--Hubbard model and the time-dependent density matrix renormalization group method to calculate 
the drag force on a trapped fermion cluster in a cloud of another fermion species with contact interaction. 
The simulation result shows that a non-trivial peak in the resistance force is observed in the high cloud density region, 
which implies a criterion of effective ways in diffusive transport in a fermion cloud. 
We compare the DMRG simulation result with a mean-field result, 
and it is suggested that some internal degrees of freedom have a crucial role in the excitation process when the cloud density is high. 
This work emphasizes the difference between the full-quantum calculation and the semiclassical calculation, which is the quantum effects, 
in slow dynamics of many-body systems bound in a fermion cloud. 
\end{abstract}

\maketitle

\section{Introduction}
Recently non-equilibrium dynamics of cold atom systems has been enthusiastically targeted, 
because cold atom systems are ideal as isolated quantum systems configured in laboratory, whose parameters can be modified dynamically \cite{review_cold_atom}. 
In cold atom systems, strength and sign of interaction between atoms can be adapted by use of Feshbach resonance \cite{review_feshbach}, and also the lattice potentials can be composed using optical lattices. 
The dynamics of quantum quench \cite{exp_quench_0, exp_quench_1, exp_quench_2, exp_quench_3, exp_quench_4} has been explored by suddenly changing the trap potential and the interaction, 
and also theoretically this dynamics has been studied 
\cite{th_quench_1, th_quench_2, th_quench_3, th_quench_4, th_quench_5, th_quench_6, th_quench_7, th_quench_8, th_quench_9, th_quench_a, th_quench_b, th_quench_c, th_quench_d, th_quench_e, th_quench_f, th_quench_g, th_quench_h, th_quench_i}. 

However, dynamics induced by a gradual change of parameters in real time has a lot more to be investigated. 
These dynamics are completely different from the quantum quench dynamics, because 
the constant change of the system parameters in time continuously causes the energy excitation and dissipation in the systems. 
In other words, these systems are not expected to relax to an equilibrium state, but they or their subsystems may reach a steady state; 
one of the goals of this work is to investigate a relaxation not to an equilibrium state but to a steady state of the whole system or only a subsystem, due to the time-dependent Hamiltonian. 

Of these dynamics, especially we focus on drag dynamics of a fermion cluster trapped by a moving trap in a fermion cloud, interacting with cluster particles by contact interaction. 
Drag dynamics is one of the basic concepts of dynamics; for example, recently spin drag dynamics has been studied 
\cite{spindrag01, spindrag02, spindrag03, spindrag04, spindrag05, spindrag06, spindrag07, spindrag08, spindrag09, spindrag10} in different situations. 
Our study aims at a detailed investigation of typical many-body drag dynamics, which is essential to the understanding of the non-equilibrium dynamics in quantum systems. 
Also, this study observes the steady state of the moving cluster in the background of the fermion cloud; 
for example, when the cluster reaches its steady state, the total energy increases linearly in time because the cluster is driven at a constant speed against the cloud, 
which means that our system as a whole is not in a steady state. 

Thus in this study we simulate the cluster drag dynamics in one-dimensional two-component Fermi systems with contact interaction. 
In this system, a cluster of fermions is forced to move by a species-dependent trap at a constant speed, interacting with a cloud of the other type of fermions by contact interaction. 
We calculate the energy of the whole system as a function of time, and evaluate the energy increase per unit time by a linear fit. 
The reason is that the total energy increase is closely related to the energy increase in the cloud, which is observable experimentally by measuring the momentum distribution of the cloud, 
since when the cluster reaches a steady state, the cloud energy increase per time equals the total energy increase per time. 
Then we compare the simulation result with a semiclassical mean-field result to clarify whether and how this dynamics is explained by a semiclassical theory. 
As a result, the characteristic peak structure in the profile of the energy increase per unit time cannot be explained by the semiclassical theory; 
internal degrees of freedom of the cluster are indispensable for the calculation in the high cloud density region. 

Our simulation is limited to one-dimensional systems, but high-dimensional versions of our system could be calculated or explored experimentally. 
Especially in cold atom systems, contact interaction between two species is realized (in this case there is no intra-species interaction because of the Fermi statistics), 
and also some kinds of spin-dependent potentials are possible. 
Our study could give some intuition for the investigation of such kind of drag dynamics in high dimensions. 

\section{Simulation}
\subsection{System setup}
\begin{figure}[htbp]
\begin{center}
\includegraphics[width=8.66truecm,clip]{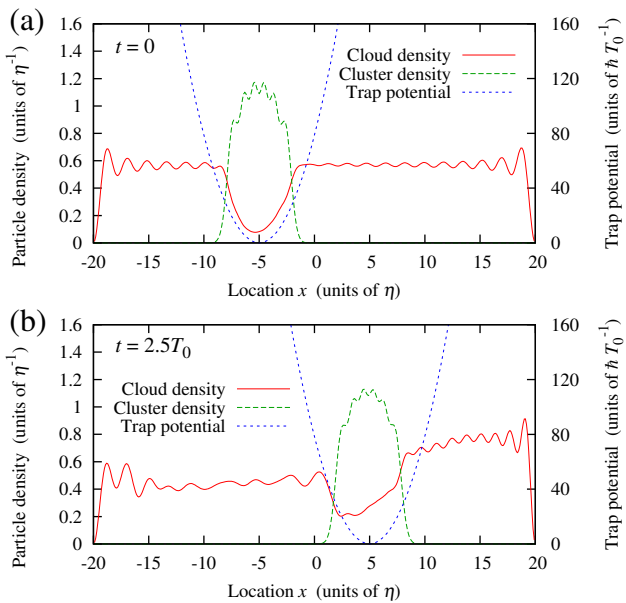}
\caption{(Color online) (a) Initial particle density and trap potential at $n=6, D=0.5\eta^{-1}, r=1, u=10\hbar T_0^{-1}\eta$. 
(b) Final particle density and trap potential at $n=6, D=0.5\eta^{-1}, r=1, u=10\hbar T_0^{-1}\eta, v=4\eta/T_0$. }
\label{trap}
\end{center}
\end{figure}

We simulate drag dynamics, and calculate the energy of the whole system in one-dimensional two-component Fermi systems. 
Initially a cluster of $n$ fermions is trapped by a harmonic potential within a cloud of the other type of free fermions (Fig.\ref{trap}(a)), where the average fermion density of the cloud is $D$ (the Fermi momentum is $\pi \hbar D$). 
The mass of a cloud fermion, which we call a background fermion below, is $m_\mathrm{back} = m_0$, 
and the mass of a cluster fermion is $m_\mathrm{cluster} = m_0 r$; the mass ratio is $m_\mathrm{cluster} / m_\mathrm{back} = r$. 
The harmonic trap potential is set as $\frac{1}{2}m_0\omega_0^2 X^2$, where $X$ is the displacement from the trap center. 
Therefore in the $r=1$ case, the frequency of the harmonic trap is $\omega_0$ (and the oscillation cycle is $T_0 = 2\pi/\omega_0$), 
and the typical width of the particle density distribution of a cluster fermion is $\eta = \sqrt{\hbar/m_0\omega_0}$. 
The interaction exists only between two fermions of the different types, and it is contact interaction which is expressed as $u\delta(x_1-x_2)$, 
in which $x_1$ and $x_2$ are the location of the two interacting fermions. 

The trap moves within the region $-5\eta \leq x \leq 5\eta$, and the whole system size is 
$L = 40\eta \gg 5\eta$; we approximate the infinite-size system by the large-size system. 
In fact, if we change the system size to $L=\infty$, an extrapolation shows that energy increase per unit time (see Sec. III) is changed by 7\% at $D=0.3\eta^{-1}$, 21\% at $D=0.6\eta^{-1}$, and 6\% at $D=0.9\eta^{-1}$, 
when the particle number of the cluster is $n = 4$, and $r = 1$. The effect of the finite system size is discussed in Sec. III. 
We set the hard-wall boundary conditions. 
The initial trap center is $x = -5\eta$ and finally moves to $x = 5\eta$. 
In this paper, we use $\eta$ and $T_0$ as the units of the system, and units of other dimensions are expressed as products of $\eta$, $T_0$ and $\hbar$; 
for example the unit of energy is $\hbar T_0^{-1}$ and the unit of power is $\hbar T_0^{-2}$. 
Therefore the independent variables in the system are $n$, $D$, $r$, $u$ and $v$, where $v$ is a trap speed as mentioned below. 
We perform calculations for conditions of $n \leq 6$ and $D \leq 1.5\eta^{-1}$ to obtain numerically exact results. 

At $t=0$ we suddenly move the trap potential by a constant speed $v$. 
Although this condition seems to be enough for the cluster to get the final speed $v$, the acceleration takes a little time. 
For faster convergence to a steady state, simultaneously we give a speed $v$ to the fermion cluster. 
Then the cluster pushes the background fermions as shown in Fig.\ref{trap}(b), while it is forced to move by the moving trap potential. 
The moving trap increases the total energy of the system $E(t)$, where we set $E(0) = 0$ (just after $v$ is given to the cluster), as a linear function of the time approximately. 
Finally the trap reaches $x = 5\eta$ in $t_F = 10\eta/v$, and then we finish the simulation. 
Fig.\ref{trap}(b) shows that the background fermions have been pushed in the positive direction. 
Later, we plot the system energy $E(t)$ as a function of time, and thus we obtain $P$, the energy increase per unit time, by a linear fit of that plot. 

\subsection{Method}
We discretize the system to adopt the one-dimensional Fermi--Hubbard model, 
and apply the time-dependent density matrix renormalization group (t-DMRG) method \cite{t-DMRG, review_DMRG2, review_DMRG3} to simulate the dynamics.
We take 399 sites numbered $-199, -198, \dots, +198, +199$ at regular intervals; the site $-50$ ($+50$) is the initial (final) location of the potential center for the fermion cluster. 
The lattice constant is $\delta x = 5\eta / 50 = 0.1\eta$, which is small enough so that the system can be treated as a continuous system: 
if we change the lattice constant into a half one, and at the same time we change the time step to preserve $[\mathrm{(time \; step)} \times \hbar^2/(2 m_0\delta x^2)]$, 
the results are changed by 0.3\% at $n=4, D=0.3\eta^{-1}$ and 4\% at $n=4, D=0.6\eta^{-1}$. 
The value of the trap potential at site $i$ is 
\begin{eqnarray}
V_{i}^\mathrm{cluster}(t) &=& \frac{1}{2}m_0\omega_0^2(x_i + 5 \eta - v t)^2 \quad (t > 0). 
\end{eqnarray}
The discretized Hamiltonian is 
\begin{eqnarray}
\hat H(t) &=& -\!\!\!\!\!\sum_{i,\sigma = \mathrm{cluster}, \mathrm{back}} \frac{\hbar^2}{2 m_\sigma \delta x^2} (\hat a^\dagger_{i,\sigma}\hat a_{i+1,\sigma} + \hat a^\dagger_{i+1,\sigma}\hat a_{i,\sigma}) \notag \\
            &+&\sum_{i}V_{i}^\mathrm{cluster}(t)\hat n_{i,\mathrm{cluster}} + \frac{u}{\delta x} \sum_{i} \hat n_{i,\mathrm{cluster}} \hat n_{i,\mathrm{back}} ,\quad\;
\end{eqnarray}
where $\hat{a}_{i,\mathrm{cluster}}$ ($\hat{a}_{i,\mathrm{back}}$) annihilates a fermion of the cluster (background) on site $i$, and $\hat{n}_{i,\sigma} \equiv \hat{a}_{i,\sigma}^\dag \hat{a}_{i,\sigma}$. 
We prepare the ground state of the system at $t = 0$ by DMRG calculation, in which the 6-7 sweeps are needed for the ground state to converge. 
Next we give the constant speed $v$ to the cluster particles as an initial state; 
starting from this state, we calculate the time evolution by that Hamiltonian with t-DMRG up to $t = t_F = 10\eta / v$. 
The time step is $2 \times 10^{-4} T_0$ and the maximum discarded eigenvalue of the reduced density matrix is below $\varepsilon = 10^{-10}$. 
In this system we apply the ``usual" second-order Suzuki-Trotter approach in spite of the time-dependent Hamiltonian, and we estimate the errors as about 1\%, changing the time step to twice or half one. 
The simulation is conducted in the following range of parameters: 
the fermion number of the cluster $n \leq 6$, 
the background fermion density (or cloud density) $D \leq 1.4\eta^{-1}$, 
the trap speed $v \sim 4\eta/T_0$, 
the mass ratio $r=$ 1 or 2, 
and the contact interaction strength $u \sim 5\hbar T_0^{-1}\eta$ or $10\hbar T_0^{-1}\eta$ 
(these values of $u$ are comparably large so that the reflection is dominant in typical collision cases). 

\section{Results}
\subsection{Time dependence and error estimation}
\begin{figure}[htbp]
\begin{center}
\includegraphics[width=8.66truecm,clip]{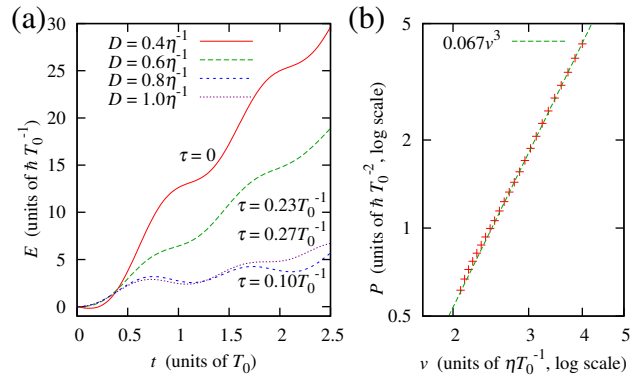}
\caption{(Color online) (a) Time dependence of the total energy $E$ at $n=6, r=1, u=10\hbar T_0^{-1}\eta, v=4\eta/T_0$ for the 4 cases: $D=0.4\eta^{-1}$, $D=0.6\eta^{-1}$, $D=0.8\eta^{-1}$, and $D=1.0\eta^{-1}$. 
(b) Trap speed dependence of total energy increase per unit time $P$ at $n=1, D=0.5\eta^{-1}, r=1, u=10\hbar T_0^{-1}\eta$ (log scale). }
\label{result_E_t_result_P_v}
\end{center}
\end{figure}

Figure \ref{result_E_t_result_P_v}(a) shows the time dependence of the total energy obtained by the DMRG simulation for various values of the cloud density $D$. 
The figure implies the linear increase in the system energy $E$ with an oscillation whose cycle is about $T_0$: the oscillation is expected to come from the motion in the trap. 
The oscillation is approximated by a exponentially dumped oscillation; 
the exponential decay constants for the 4 cases in Fig.\ref{result_E_t_result_P_v}(a) are evaluated at 
about 0 (no convergence during the simulation time), 0.23$T_0^{-1}$, 0.10$T_0^{-1}$, and 0.27$T_0^{-1}$ respectively. 
This fact suggests that, in these three cases, the cluster loses the effect of the initial conditions. 
Therefore we evaluate $P$, the energy increase per unit time, by a linear fit of this plot neglecting the oscillation, so that we extract the asymptotic behavior in the longer period of time: 
the linear fit is conducted over one or two cycles of internal motions in the trap ($\chi^2$ fit in the region $0.5<t/ T_0\leq 2.5$ for $r=1$ and $2.5-\sqrt{2}<t/ T_0\leq 2.5$ for $r=2$). 
In other words, $P$, which corresponds to the power of the trap motion, is expected to give approximate information of the steady state of the cluster.
The resistance force $F$ against the trap is calculated by the relation $Fv = P$; $P$ is proportional to $F$ when the trap speed $v$ is fixed. 

The errors of $P$ come from the DMRG simulation ($<5\%$), the linear-fit, the discretization effect ($\sim5\%$), and the finite size effect. Here the errors of the linear fit are of the same order as those of the DMRG simulation. 
This is because the linear fit errors mainly come from the dumped oscillation (comparably weak during a cycle) and from the uncertainty of the oscillation cycle, 
which is assumed as $T_0$ for $r=1$ and $\sqrt{2} T_0$ for $r=2$, and in our system, the oscillation amplitude is of the same or smaller order than the one-cycle difference of the energy. 

\begin{figure}[htbp]
\begin{center}
\includegraphics[width=8.66truecm,clip]{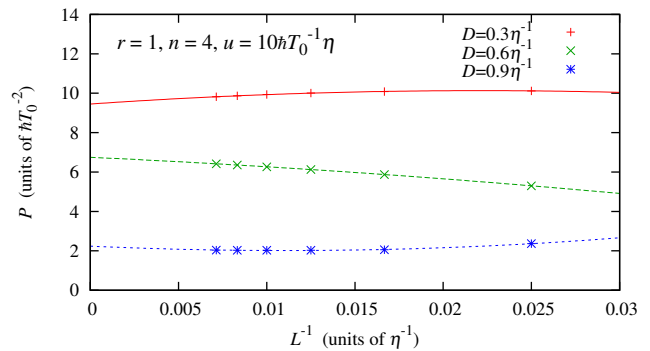}
\caption{(Color online) Effect of the finite system size: total energy increase per unit time $P$ as a function of the inverse of the system size $L^{-1}$ at $n=4, r=1, u=10\hbar T_0^{-1}\eta$.  }
\label{syssize}
\end{center}
\end{figure}

In our system, the finite size effect is dominant and it determines the order of the total errors. 
The finite size effect is shown in Fig.\ref{syssize}, where the site number is changed so that the lattice constant is fixed to $\delta x = 0.1\eta$. 
The system size dependence is extrapolated by quadratic functions of $L^{-1}$. 
The finite size effect is estimated to be $|P(L=40\eta)-P(L=\infty)|/P(L=\infty)$, where $P(L=\infty)$ is the extrapolated data in the limit of $L^{-1} \to 0$. 
The ratio is 7\% at $D=0.3\eta^{-1}$, 21\% at $D=0.6\eta^{-1}$, and 6\% at $D=0.9\eta^{-1}$; thus the finite size effect is of the order of 20\% or smaller. 
Therefore the total errors of the slope $P$ are estimated to be about 20\%. 

\subsection{Velocity and interaction dependence}
We focus on the dependence of the value of $P$ on the simulation parameters in the following. 
To see the velocity dependence of $P$, we plot $P$ against $v$ in the case of $n=1, r=1$ in Fig.\ref{result_E_t_result_P_v}(b). 
The figure demonstrates the relation $P \propto v^3$, so that the reaction force to the trap is proportional to $v^2$; this suggests the existence of the inertial resistance. 
The inertial resistance is physically expected, because this system is one-dimensional and a cluster particle gives its momentum to a background fermion in a single collision, 
whose rate is proportional to $v$. 

\begin{figure}[htbp]
\begin{center}
\includegraphics[width=8.66truecm,clip]{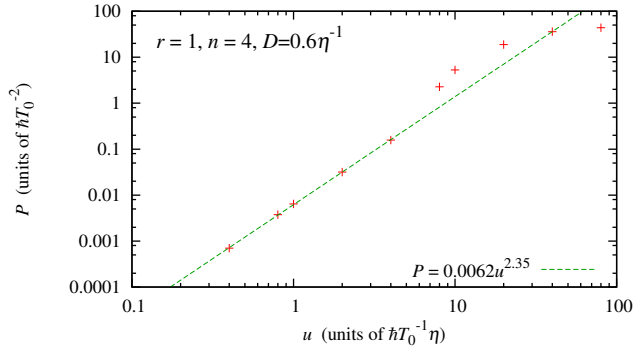}
\caption{(Color online) Energy increase per unit time $P$ versus interaction strength $u$ at $r=1$, $n=4$, $D=0.6$, and $v=4\eta/T_0$. }
\label{result_P_u}
\end{center}
\end{figure}

Then, we show the interaction dependence of $P$ in Fig.\ref{result_P_u} at $n=4, D=0.6$. 
This figure indicates that $P$ is proportional to $u^{2.35}$ in the small $u$ region, but seems to be saturated in the limit of $u\rightarrow \infty$: 
even if $u$ is infinite, $P$ remains finite because infinite $u$ corresponds to a moving hard wall. 
Although we cannot explain the exponent, the interaction strength has little effect on other than the peak height at $u \sim 10$ as discussed below. 
In our simulation, we do not use attractive interaction to avoid any possibility of initial bound state, 
but we believe that the sign of the interaction has no effect on $P$ if there is no initial bound state, 
because the sign inversion of the interaction only changes the phase sign of scattered wavefunctions in scattering processes, and then makes no change in observables. 

\begin{figure}[htbp]
\begin{center}
\includegraphics[width=8.66truecm,clip]{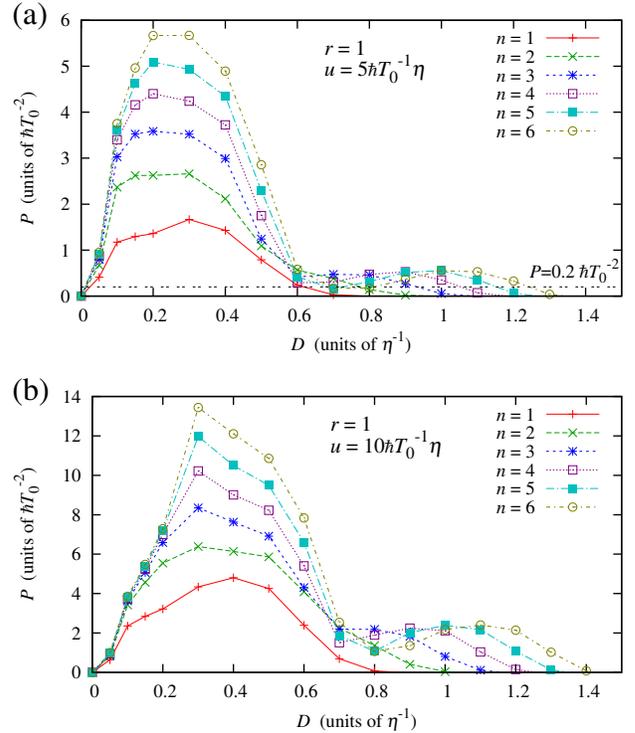}
\caption{(Color online) Energy increase per unit time $P$ as a function of cloud density $D$ for $1 \leq n \leq 6$ and $0 \leq D \leq 1.4\eta^{-1}$ at $r=1$ and $v=4\eta/T_0$. 
(a) is $u=5\hbar T_0^{-1}\eta$ case, and (b) is $u=10\hbar T_0^{-1}\eta$ case. }
\label{result_P_D_r1}
\end{center}
\end{figure}

\subsection{Cloud density and other dependence}
Next we investigate how $P$ depends on the cloud density and the number of the cluster particles. 
Figures \ref{result_P_D_r1}(a) and (b) show the energy increase per unit time $P$ for $0.3\eta^{-1} \leq D \leq 1.5\eta^{-1}$ 
in the cases of $u=5\hbar T_0^{-1}\eta$ and $u=10\hbar T_0^{-1}\eta$, respectively. 
In the figures, $D$ goes to zero exponentially in large $D$; no resistivity is observed in this region. 
Also, the figures exhibit a single-peak structure for the cases of $n=1$ or $n=2$, and a double-peak structure for the cases of $n \geq 3$. 
In the large $D$ region, the decay curves of $3\leq n \leq 6$ have very similar shapes at regular intervals, and the distance between the neighbors is about $0.10\eta^{-1}$ 
(i.e., the values of $D \eta$ when $P$ is $0.2 \hbar T_0^{-2}$ after the second peaks in Fig.\ref{result_P_D_r1}(a) are 0.93 at $n=3$, 1.05 at $n=4$, 1.15 at $n=5$ and 1.24 at $n=6$). 
This suggests that the second peaks are at regular intervals of about $0.1\eta^{-1}$. 
These characteristics does not strongly depend on interaction strength $u$. 

The important result obtained from the figures \ref{result_P_D_r1}(a) and (b) is that $P < nP_{n=1}$ ( $P > nP_{n=1}$ ) holds for small (large) $D$, where $P_{n=1}$ is $P$ at $n=1$.
In other words, if one wants to reduce the energy increase in moving $n$ fermions in a fermion background, 
one should trap the fermions in a single trap in the small $D$ case, but in $n$ traps independently in the large $D$ case; 
this is a criterion of ways for energy saving in diffusive transport in a fermion background. 
In the following we mainly focus on the peak structure, especially the peak location. 

\begin{figure}[htbp]
\begin{center}
\includegraphics[width=8.66truecm,clip]{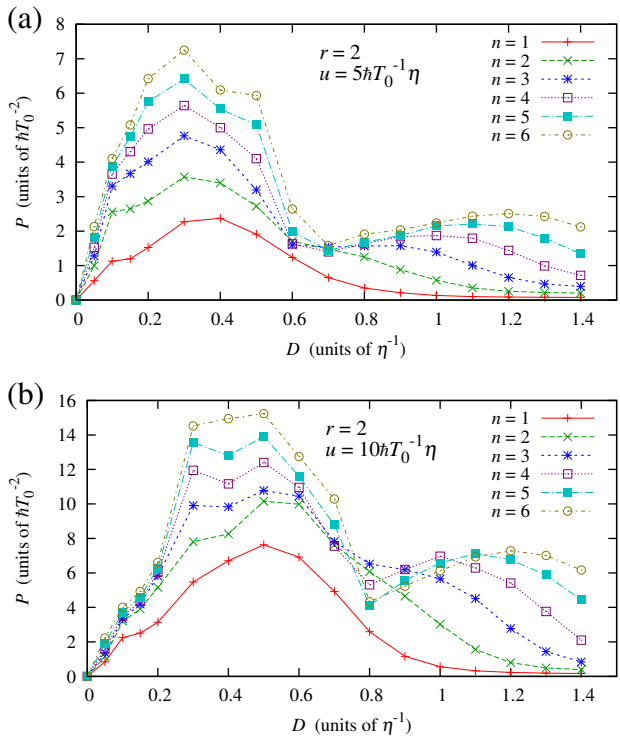}
\caption{(Color online) Energy increase per unit time $P$ as a function of cloud density $D$ for $1 \leq n \leq 6$ and $0 \leq D \leq 1.4\eta^{-1}$ at $r=2$ and $v=4\eta/T_0$. 
(a) is $u=5\hbar T_0^{-1}\eta$ case, and (b) is $u=10\hbar T_0^{-1}\eta$ case. }
\label{result_P_D_r2}
\end{center}
\end{figure}

Let us now explore the dependence of $P$ on the mass ratio $r$. Figures \ref{result_P_D_r2}(a) and (b) show $P$ in the same parameters as in Fig.\ref{result_P_D_r1}, except for $r=2$. 
Except that the height of the second peak differs between different values of $n$, these plots have the same features as in Fig.\ref{result_P_D_r1}. 
To extract the locations of the second peaks, we interpolate the data close to the second peak as a quadratic curve, and obtain the average distance $0.1\eta^{-1}$
(i.e., the estimated values of $D \eta$ at the peak in Fig.\ref{result_P_D_r2}(a) are 0.86 at $n=3$, 0.98 at $n=4$, 1.09 at $n=5$ and 1.20 at $n=6$); 
the distance between the neighboring second peaks has values similar to the $r=1$ case. 

\subsection{Cluster energy}
\begin{figure}[htbp]
\begin{center}
\includegraphics[width=8.66truecm,clip]{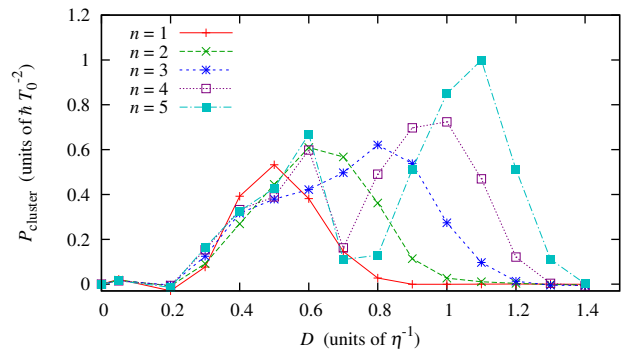}
\caption{(Color online) Increase of cluster energy per unit time $P_\mathrm{cluster}$ versus cloud density $D$ for $1 \leq n \leq 5$ and $0 \leq D \leq 1.4\eta^{-1}$ at $r=1, u=10\hbar T_0^{-1}\eta, v=4\eta/T_0$. }
\label{result_Pcluster_D}
\end{center}
\end{figure}

Here we examine the energy of the cluster, the expectation value of the cluster Hamiltonian
\begin{eqnarray}
\hat H_\mathrm{cluster}(t) = &-&\!\!\!\!\!\sum_{i,\sigma = \mathrm{cluster}} \frac{\hbar^2}{2 m_\sigma \delta x^2} (\hat a^\dagger_{i,\sigma}\hat a_{i+1,\sigma} + \hat a^\dagger_{i+1,\sigma}\hat a_{i,\sigma})\notag\\
            &+&\sum_{i}V_{i}^\mathrm{cluster}(t)\hat n_{i,\mathrm{cluster}}\;, 
\end{eqnarray}
which is calculated in the case of $r=1, u=10\hbar T_0^{-1}\eta$. 
We plot the cluster-energy increase per unit time in Fig.\ref{result_Pcluster_D}, which is evaluated by a linear fit of the expectation values; 
the linear fit is done by averaging two lines which come in contact with two consecutive cycles of the oscillation curve of the cluster energy, 
because in this case a $\chi^2$ linear fit under large oscillation compared to the slope gives large estimation errors. 
In the figure, a peak (peaks) are observed in the case of $n=1,2$ ($n>2$). 
In the peak region, the energy of the cluster increases over time. 
Therefore the cluster does not completely reach a steady state in the finite simulation time, since the cluster energy should not change in a steady state. 
On the other hand, we observe that the system reaches a steady state out of the peak region in our simulation time. 
The values of $D$ giving the peak-structure are a bit larger than those in Fig.\ref{result_P_D_r1}(b), so the excessive energy flux into the cluster corresponds to the decreasing $P$. 
In other words, the steady state assumed becomes the farthest from the initial state just after the peak region of $P$. 
Nevertheless, the total energy increase in Figs. \ref{result_P_D_r1}(a) and (b) 
approximately represents the energy increase per unit time after the steady state (with no increase in the cluster energy) has been established, 
because the cluster-energy increase is much smaller than the total energy increase. 

\section{Discussion}
We compare \naive mean-field results with the DMRG results, in order to investigate the many-body effects in the system. 
Under a mean-field approximation, the background particles move in the potential created by the average interaction with the cluster, 
which is calculated by the density distribution of the cluster particles of the ground state (e.g., a Gaussian function at $n=1$). 
The potential moves by velocity $v$, pushes the cloud, and excites the system. 
The mean-field Hamiltonian of the background particles is 
\begin{eqnarray}
H_\mathrm{MF}(t) &=& -\frac{\hbar^2}{2 m_0}\frac{\partial^2}{\partial x^2} + V_\mathrm{MF}(x-v t), \\
V_\mathrm{MF}(x) &=& u\sum_{k=0}^{n-1}|\psi_k(x)|^2 ,\\
\psi_k(x)        &=& C_k H_k\left(\sqrt{\frac{r m_0 \omega_0}{\hbar}}x\right) \exp \left(-\frac{r m_0 \omega_0}{2\hbar}x^2\right), 
\end{eqnarray}
where $H_k$ are the Hermite polynomials and $C_k$ are normalization constants; 
$\psi_k(x)$ are the wavefunctions of the cluster particles trapped by a harmonic trap. 

The energy increase per unit time under this Hamiltonian is calculated by solving the scattering problem for the potential $V_\mathrm{MF}(x)$. 
On the coordinates fixed to the trap, the reflectance $R(p)$ can be calculated as a function of incident momentum $p$. 
Within the time $dt$, the number of reflected particles whose momenta are between $p$ and $p+dp$ is 
\begin{equation}
 \frac{(|p|/m_0)dt \cdot dp}{2\pi\hbar} R(p) = dt \frac{|p|R(p)dp}{2\pi\hbar m_0} . 
\end{equation}
Therefore the energy increase per unit time can be computed as 
\begin{eqnarray}
P_\mathrm{MF} &=& \int_{-\pi \hbar D-m_0 v}^{\pi \hbar D-m_0 v}\frac{|p|R(p)dp}{2\pi\hbar m_0} \frac{(-p +m_0 v)^2 - (p +m_0 v)^2}{2 m_0} \nonumber \\
&=& \frac{v}{\pi\hbar m_0}\int_{|\pi \hbar D-m_0 v|}^{\pi \hbar D+m_0 v}p^2 R(p) dp .
\end{eqnarray}
We calculate $R(p)$ and $P_\mathrm{MF}$ numerically, in the case of $r=1, u=10\hbar T_0^{-1}\eta, v=4\eta/T_0$. 

\begin{figure}[htbp]
\begin{center}
\includegraphics[width=8.66truecm,clip]{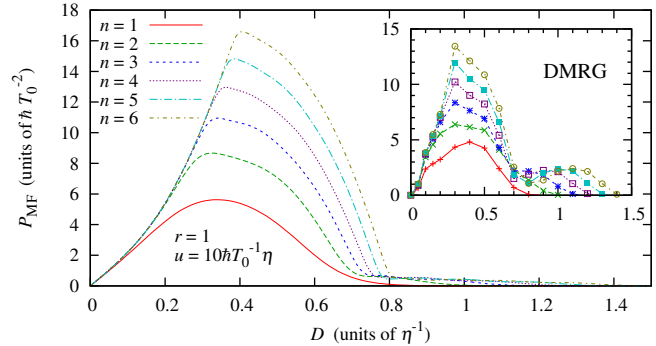}
\caption{(Color online) The mean-field results compared to the DMRG results: energy increase per unit time obtained by the mean-field approximation $P_\mathrm{MF}$ for $1 \leq n \leq 6$ at $r=1, u=10\hbar T_0^{-1}\eta, v=4\eta/T_0$.}
\label{result_meanfield}
\end{center}
\end{figure}

The mean-field results are compared with the DMRG results in Fig. \ref{result_meanfield}. 
We find that they have the same tendency in the small $D$ region, although they are slightly different in their scale of $D$ and $P$. 
However, they disagree in the large $D$ region: the mean-field results have small plateaus, instead of peaks which are observed in the DMRG results. 
This is partially because the mean-field potential does not strongly scatter the background fermions with larger momenta than the Fermi momentum of the cluster;
if the Fermi momentum of the background is much larger than that of the cluster, the moving mean-field potential has little effect. 
Thus, while the mean-field theory can explain this dynamics for small $D$ with a slight transformation in scale, 
it fails to explain the second peaks in the large $D$ region. 
Therefore the many-body effects such as the internal degrees of freedom of the cluster have a crucial role in forming the peak structure in the large $D$ region; 
the possible model for this system, which is one of the future works, has to contain its many-body effects.

\section{Conclusion}
In summary, using the time-dependent density matrix renormalization group method and the Fermi--Hubbard model, 
we have calculated the drag dynamics of several fermions in a fermion cloud in one-dimensional continuous systems. 
We have obtained the steady energy increase per unit time as a function of the particle number of the cluster $n$, the cloud density, 
the mass ratio between fermions, the interaction strength, and the trap speed. 
We have discovered the emergence of a double-peak structure; one is in the low cloud density region, and the other is in the high cloud density region. 
We have revealed that when one wants to reduce the system excitation in moving a fermion cluster in a fermion cloud, 
one should move the cluster packed together if the cloud density is low, 
but one should move the fermions in separate $n$ traps if the cloud density is higher. 
We have introduced a mean-field approximation for the system to estimate whether and how this dynamics can be explained by semiclassical models. 
We have elucidated that the drag dynamics in the high density fermion cloud cannot be explained by our mean-field model, 
while it can be reproduced with a small scale transformation in the low cloud density; 
we have emphasized the quantum effects in the drag dynamics of a cluster bound in a fermion cloud. 

\subsection*{Acknowledgments}
The work of J. O. was supported by a JSPS Research Fellowship for Young Scientists.
This work was in part supported by JSPS KAKENHI Grant Numbers 26870284 (M. T.), 25400366, and 15H05855 (N. K.).
Part of numerical computation in this work was carried out at the Supercomputer Center, ISSP, University of Tokyo.

\end{document}